\documentclass[pre,twocolumn,amsmath,showpacs]{revtex4}
\usepackage{graphicx}
\begin{document}
\title{Roughening transition driven by a binary spreading process}
\author{Haye Hinrichsen}
\affiliation{Theoretische Physik, Fachbereich 8, Universit{\"a}t
	     Wuppertal, 42097 Wuppertal, Germany}

\begin{abstract}
We introduce a solid-on-solid growth process which evolves by random deposition of dimers, surface diffusion, and evaporation of monomers from the edges of plateaus. It is shown that the model exhibits a robust transition from a smooth to a rough phase. The roughening transition is driven by an absorbing phase transition at the bottom layer of the interface, which displays the same type of critical behavior as the pair contact process with diffusion $2A\to 3A$, $2A\to\emptyset$.
\end{abstract}

\pacs{05.70.Ln, 64.60.Ak, 64.60.Ht}

\maketitle
\parskip 1mm 

\section{Introduction}
%
The fabrication of thin films by deposition and evaporation of particles plays an important role in technological applications such as molecular beam epitaxy. Theoretical studies of growing films are usually based on simplified models of interfaces evolving by certain stochastic dynamic rules. Even simple solid-on-solid models may display very interesting features such as kinetic roughening, scaling, self-similarity, and fractal properties~\cite{Meakin,KrugSpohn,Dietrich,Barabasi,Krug}. 
\begin{figure*} 
\includegraphics[width=160mm]{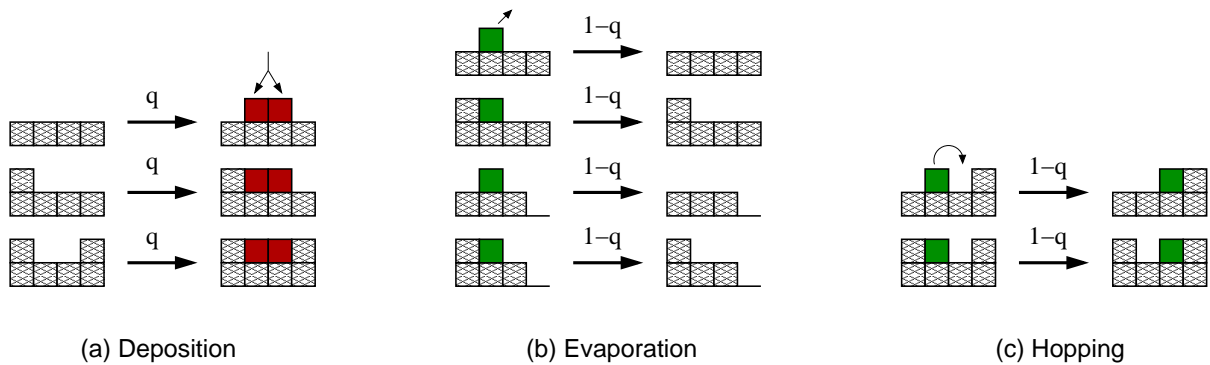}
\caption{\label{FIGRULES}
List of all possible rearrangements of the interface at the sites $i-1,i,i+1,i+2$ according to the dynamic rules~(\ref{Deposition})-(\ref{RSOS}): {\bf (a)} Deposition of a dimer at the sites $i$ and $i+1$. {\bf (b)} Evaporation of a monomer from the right edge of a terrace at site $i$. {\bf (c)}~Hopping of a monomer between two adjacent islands, moving the hole between the islands in opposite direction. In all cases the spatially reflected rules (not shown here) have to be included as well.
}
\end{figure*}

Particularly fascinating are models with a roughening transition from a smooth phase, where the width of the interface is stationary and finite, to a rough phase, where the width grows continuously. Roughening transitions may rely on different physical mechanisms. For example, Kardar-Parisi-Zhang (KPZ) growth processes are known to exhibit a roughening transition in $d>2$ dimensions, which is caused by a competition of surface tension and the nonlinear term in the equation of motion. Another class of roughening transitions, on which we will focus in the present work, occurs in certain deposition-evaporation processes with the special feature that atoms cannot desorb from the middle of plateaus. In these models a completed layer, once formed, is stable and cannot evaporate again. Thus, if the growth rate is sufficiently small, the interface is smooth and fluctuates close to the lowest completed layer (the so-called bottom layer of the interface), while for large deposition rate the growing islands coalesce so that the interface propagates and roughens continuously.

A very simple model, which exhibits such a roughening transition, has been introduced some time ago by Alon {\it et al.}~\cite{Alon1}. The model is defined as a restricted solid-on-solid growth process with the special property that atoms can only evaporate only at the edges of plateaus. It turned out that the  roughening transition in this model is driven by a directed percolation (DP) process  at the bottom layer. DP is the standard universality class of phase transitions from a fluctuating phase into an absorbing state and occurs in unary reaction diffusion processes of the form $A\to 2A$, $A\to \emptyset$~\cite{Hinrichsen00a}. If the decay $A\to\emptyset$ dominates the system eventually enters an absorbing state without particles from where it cannot escape, while in the active phase the reaction $A\to 2A$ is strong enough to sustain a stationary density of particles. As shown in Ref.~\cite{Alon1} the sites at the  bottom layer of the interface can be interpreted as particles of a DP process. Using this relation the pinned phase, where the interface is smooth, corresponds to the active phase of DP. However, if the growth rate is increased above a certain critical threshold the DP process eventually enters the absorbing state, meaning that the bottom layer propagates by one step. 

The relation to DP allows one to predict some of the critical properties at the roughening transition. Since the density of sites at the bottom layer~$n_0$ should scale in the same way as the particle density in DP, it is expected to decay at criticality as
\begin{equation}
\label{Decay}
n_0 \sim t^{-\delta} \,,
\end{equation}
where $\delta =\beta/\nu_\perp$ is one of the universal critical exponent of DP. Extending these arguments it has been shown that even the critical behavior at the first few layers above the bottom layer can be described in terms of a unidirectionally coupled hierarchy of DP processes~\cite{Coupled}. Roughly speaking each layer is associated with a separate DP process which is coupled to the dynamics at the layer below, while effective  couplings in opposite direction turn out to be irrelevant in the renormalization group sense.

More recently it became apparent that other universality classes of phase transitions into absorbing states can also be used to generate a roughening transition in an appropriately defined growth model. For example, the so-called parity-conserving (PC) class~\cite{PC}, which is most prominently represented by branching-annihilating random walks with two offspring $A\to 3A$, $2A\to\emptyset$, has been shown to generate a roughening transition in a simple model for dimer adsorption and desorption~\cite{Dimer}. Therefore, the question arises whether it is possible to relate {\em any} universality class of absorbing phase transitions to a corresponding class of roughening transitions.

As a complete classification scheme of absorbing phase transitions is not yet available, it is an important task to search for other yet unknown types of transitions. In this context one of the most promising and controversially debated candidates is the pair contact process with diffusion (PCPD)
\begin{equation}
\label{PCPD}
2A \to 3A, \quad 2A \to \emptyset \,.
\end{equation}
In contrast to  DP and PC transitions, where individual particles can generate offspring, the PCPD is a {\em binary}  spreading process, i.e., two particles have to meet in order to create a new one. Although there is a still ongoing debate to what extent the PCPD represents a genuine independent universality class, there seems to be a general consensus that it displays a novel type of phase transition that has not been seen in other models~\cite{Grassberger82,HowardTauber97,Carlon01,Hinrichsen01,Odor00,CyclicPaper,Odor01,PCPCPD,Odor02,Odor02b,HenkelHinrichsen01,NohPark02,ParkKimNew,Hinrichsen02,Chate02,Odor02c}. In the numerically accessible temporal regime the transition is characterized by a fairly reproducible set of effective critical exponents. Very recently Kockelkoren and Chat\'e~\cite{Chate02} carried out extensive simulations, reporting the estimates
\begin{equation}
\label{ChateExpon}
\beta=0.37(2)\,, \quad \delta=\frac{\beta}{\nu_\parallel}=0.200(5)\,, 
\quad z=\frac{\nu_\parallel}{\nu_\perp}=1.70(5)\,.
\end{equation}
The purpose of this paper is to introduce a solid-on-solid growth process which exhibits a roughening transition driven by a {\em binary} spreading process at the bottom layer. Focusing on the one-dimensional case, where fluctuation effects are most pronounced, we demonstrate that the pair contact process with diffusion can indeed be used to generate a roughening transition in an appropriately defined model for interface growth. 

\section{Definition of the model}
%
The dynamic rules of the model involve three different physical processes, namely, deposition of dimers, surface diffusion, and evaporation of diffusing monomers. The dimers are deposited horizontally on pairs of sites at equal height, leading to the formation of islands. These islands are stable in the interior but slightly unstable at the edges, where monomers are released at a certain rate. These monomers then diffuse on the surface until they either attach to another island or evaporate back into the gas phase. In what follows we consider the limit of a very high rate for evaporation. In this limit a released monomer is most likely to evaporate unless it immediately attaches to an {\em adjacent} island at the next site, effectively moving the hole between the two islands in opposite direction. 

In more technical terms, the model is defined on a one-dimensional lattice with periodic boundary conditions. Each site $i$ is associated with an integer height variable~$h_i$. The model evolves by random-sequential updates, i.e., a site $i$ is randomly selected and one of the following moves is carried out (see Fig.~\ref{FIGRULES}):
\begin{itemize}
\item[{\bf (a)}] With probability $q$ a dimer is deposited horizontally provided that the two supporting sites have the same height:
\begin{equation} 
\label{Deposition}
\begin{split}
&\text{ If } h_i=h_{i+1} \text{ then }\\
&\qquad  h_i\to h_i+1; \;
h_{i+1}\to h_{i+1}+1 
\end{split}
\end{equation}
\item[{\bf (b)}] With probability $1-q$ a monomer is released at the edge of a terrace. If there is no adjacent terrace, the released monomer evaporates immediately:
\begin{equation} 
\label{Evaporation}
\begin{split}
&\text{ If }
h_i > h_{i+1} \geq h_{i+2} \text{ then } \\
&\qquad h_i\to h_i-1 \\[2mm]
&\text{ If } h_{i+1} > h_{i} \geq h_{i-1} \text{ then }\\
&\qquad h_{i+1}\to h_{i+1}-1
\end{split} 
\end{equation}
\item[{\bf (c)}] Otherwise the released monomer attaches to the opposite edge of the adjacent terrace:
\begin{equation} 
\label{Hopping}
\begin{split}
&\text{ If } h_i > h_{i+1} < h_{i+2}  \text{ then }\\
&\qquad h_i\to h_i-1;\; h_{i+1}\to h_{i+1}+1 \\[2mm]
&\text{ If } h_{i+1} > h_{i}< h_{i-1} \text{ then }\\
&\qquad  h_{i+1}\to h_{i+1}-1;\; h_{i}\to h_{i}+1
\end{split} 
\end{equation}
\end{itemize}
In addition, an attempted move is abandoned if the resulting configuration would violate the restricted solid-on-solid (RSOS) condition 
\begin{equation}
\label{RSOS}
|h_i - h_{i \pm 1}| \leq 1\,.
\end{equation}
The RSOS condition introduces an effective surface tension and imposes an infinitely large Ehrlich-Schwoebel barrier, i.e., particles cannot diffuse across the edges of terraces. Each attempted update corresponds to a time increment of $\Delta t=1/L$. Note that the dynamic rules are translationally invariant in horizontal as well as in vertical direction.  

\section{Phenomenological properties and relation to the PCPD}
%
%
%
%
\begin{figure} 
\includegraphics[width=75mm]{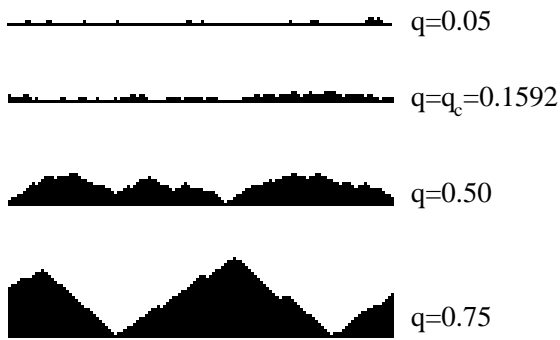}
\caption{\label{FIGDEMO}
Typical interface configurations for various values of the growth rate~$q$.
}
\end{figure}

The growth process defined above has the following phenomenological properties (see Fig.~\ref{FIGDEMO}). For small values of~$q$ occasionally deposited dimers will quickly desorb from the surface by evaporation at the edges so that the interface is smooth and fluctuates close to the bottom layer. As $q$ is increased the islands grow and coalesce until their average size diverges at a certain critical point $q_c$. Above the critical point more and more islands are deposited on top of each other so that the interface roughens continuously. However, the interface remains pinned at solitary sites at the bottom layer, leading to the formation of pyramidial structures. These pyramids fluctuate so that the pinning sites slowly diffuse until they merge or annihilate. In contrast to the dimer model introduced in Ref.~\cite{Dimer} there is no evidence for a facetting transition at $q=1/2$.

In order to explain how the growth model is related to the PCPD, let us consider the dynamic processes at the bottom layer (the spontaneously selected lowest height level of the interface). Interpreting sites at the bottom layer as particles $A$ of a spreading process, the dynamic rules listed in Fig.~\ref{FIGRULES} can be associated with certain reactions of the particles. For example, the deposition of a dimer corresponds to pairwise annihilation $2A\to\emptyset$, while evaporation of a monomer can be viewed as the creation of a particle $A$. However, in the present model atoms can only evaporate at the edge of a terrace followed by two bottom layer sites, hence rule {\bf (b)} in Fig.~\ref{FIGRULES} has to be interpreted as a fission process $2A\to 3A$. Otherwise, if there is only one such bottom layer site next to the edge, rule~{\bf (c)} applies which corresponds to a random walk of a single particle $A$. Thus we may interpret the dynamics at the bottom layer as a reaction-diffusion process
\begin{equation}
\begin{split}
\text{\bf (a)} \quad & AA \to \emptyset\emptyset \\
\text{\bf (b)} \quad & AA\emptyset\, /\, \emptyset AA \to AAA \\
\text{\bf (c)} \quad & \emptyset A \leftrightarrow A \emptyset
\end{split}
\end{equation}
which resembles the dynamic rules of the pair contact process with diffusion in a fermionic realization. Clearly, the relation between the growth model and the PCPD is not rigorous, especially because of deposition-evaporation processes on top of islands. However, as in the case of DP- and PC-related growth models~\cite{Alon1,Dimer}, this correspondence is expected to be valid in the renormalization group sense and therefore determining the asymptotic critical behavior.

\section{Critical properties at the roughening transition}
%
%
%
\begin{figure} 
\includegraphics[width=80mm]{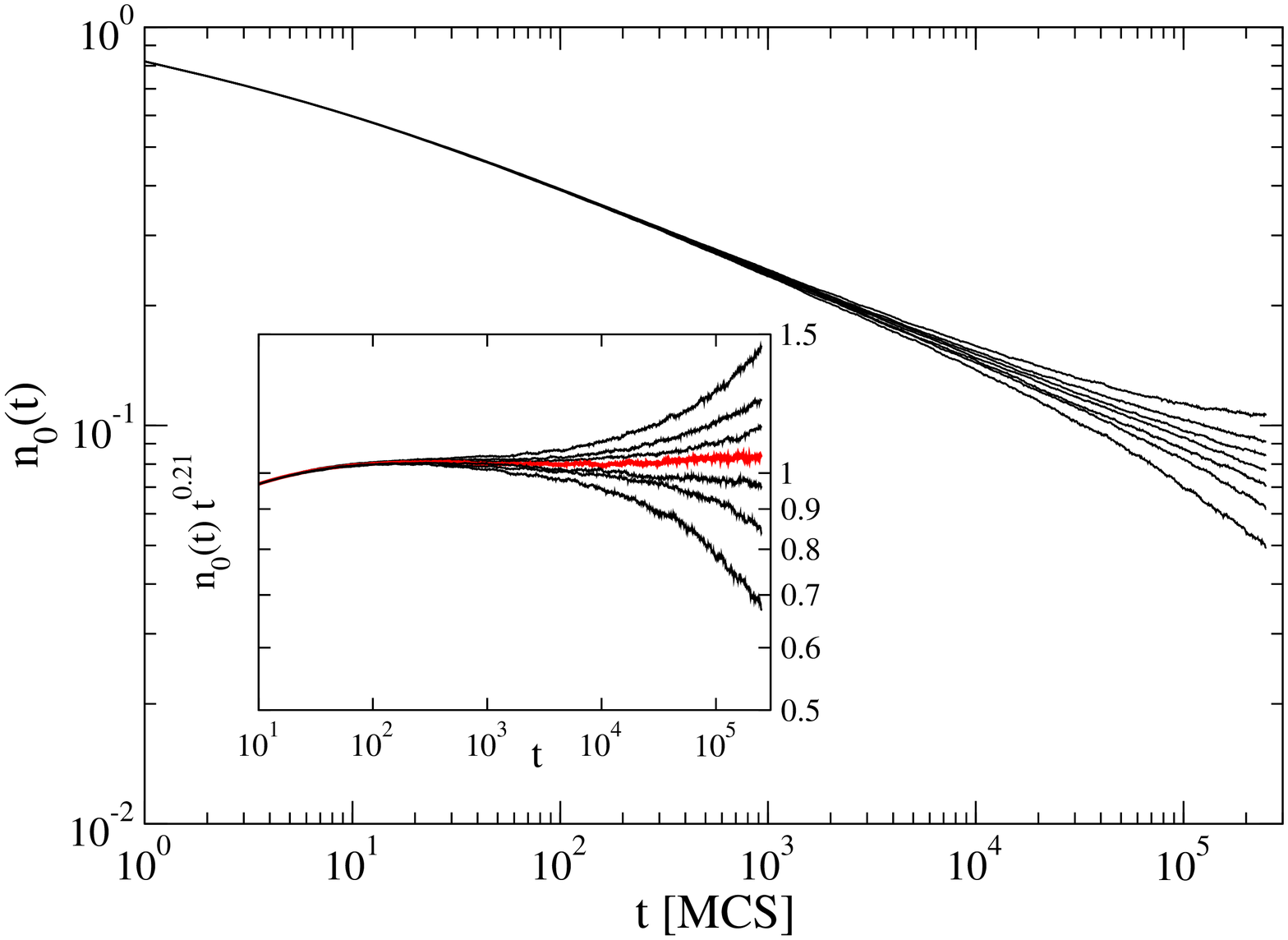}
\caption{\label{FIGDECAY}
Density of sites at the lowest height (bottom layer) as a function of time for $q=0.15900$, $0.15910$, $0.15915$, $0.15920$, $0.15925$, $0.15930$, and $0.15940$ averaged over up to $1000$ independent runs. The inset shows the same data multiplied with $t^\delta$, where $\delta=0.21(2)$ is the estimated exponent.
}
\end{figure}
%
%
%
\subsection{Scaling at the critical point}

In order to confirm that the roughening transition in the present model is driven by a binary spreading process, we show that the density of sites at the bottom layer~$n_0$ displays the same type of decay as the density of particles in the PCPD. To this end we perform numerical simulations starting with an initially flat interface. Measuring $n_0$ for various values of $q$, the critical threshold $q_c$, where the density decays approximately as a power-law, is estimated by
\begin{equation}
q_c = 0.15920(5)\,.
\end{equation}
The corresponding exponent
\begin{equation}
\label{delta}
\delta = \frac{\beta}{\nu_\parallel} = 0.21(2)
\end{equation}
is indeed in good agreement with the expected PCPD exponent~(\ref{ChateExpon}).
However, the plotted lines are still slightly curved, a problem which also occurs in many other binary spreading processes. 

\begin{figure} 
\includegraphics[width=70mm]{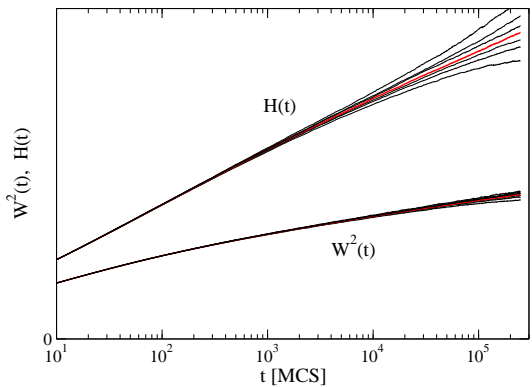}
\caption{\label{FIGWH}
Mean height $H(t)$ and squared width $W^2(t)$ as functions of time in a semilogarithmic plot for the same parameters as in Fig.~\ref{FIGDECAY}.
}
\end{figure}

Fig.\ref{FIGWH} shows the the average height
\begin{equation}
H(t) = \frac{1}{L} \sum_{i=1}^L h_i(t)
\end{equation}
and the squared width
\begin{equation}
W^2(t) = \frac{1}{L} \sum_{i=1}^L \bigl(h_i(t)-H(t)\bigr)^2
\end{equation}
as functions of time. In previously investigated DP- and PC-driven roughening transitions these quantities were found to increase {\em logarithmically} with time at the critical point. In the present case the average height shows an almost logarithmic increase $H(t) \simeq 0.19 + 0.13 \ln t$, whereas the squared width $W^2(t)$ does not display a convincing logarithmic law, at least within the numerically accessible range.
%
%

\subsection{Finite-size scaling}

%
%
\begin{figure} 
\includegraphics[width=90mm]{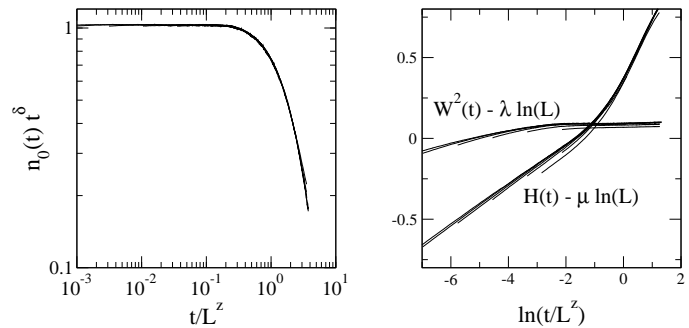}
\caption{\label{FIGFS}
Finite size scaling. Data collapses according to the scaling forms~(\ref{Form1}), (\ref{Form2}) and~(\ref{Form3}) for system sizes L=32, 64,  128, $\ldots$, 1024 averaged over $15000$ independent runs. Data points for $t<50$ are eliminated in order to suppress deviations due to initial transients. Moreover, the curves are cut for $t>4\times L^{1.75}$, from where on the interface may become pinned at a single bottom layer site, leading to additional deviations for large~$t$.
}
\end{figure}
In order to estimate the dynamic exponent $z$ we performed finite-size simulations. According to the standard scaling theory the bottom layer density $n_0$ is expected to obey the finite-size scaling form
\begin{equation}
\label{Form1}
n_0(L,t) = t^{-\delta} \, f(t/L^z)\,,
\end{equation}
where $f(z)$ is a universal scaling function. This scaling form can be used to generate a data collapse by potting $n_0(L,t)t^\delta$ versus $t/L^z$, as shown in the left panel of Fig.~\ref{FIGFS}. Using the previous estimate~(\ref{delta}) and tuning $z$ we obtain an excellent data collapse for 
\begin{equation}
z=1.75(2)\,,
\end{equation}
which is in agreement with the value in Eq.~(\ref{ChateExpon}) for the PCPD. As in DP- and PC-driven roughening transitions, we expect this exponent to be valid not only at the bottom layer but for the growth process as a whole. Therefore, the dynamic exponent $z$ should also determine the finite-size scaling properties of the squared width $W^2(L,t)$ and the average height $H(L,t)$. In the case of DP-driven roughening transitions these quantities were found to obey the scaling forms~\cite{Alon2} 
\begin{eqnarray}
\label{Form2}
\exp\bigl(W^2(L,t)\bigr) &=& L^\lambda F_1(t/L^z) \,,\\
\label{Form3}
\exp\bigl(H(L,t)  \bigr) &=& L^\mu     F_2(t/L^z) \,
\end{eqnarray}
so that it is natural to assume that the same forms hold in the present case. In fact, plotting $W^2(L,t)-\lambda\ln L$ and $H(L,t)-\mu \ln L$ versus $\ln(t/L^z)$ both quantities show a reasonable data collapse for
\begin{equation}
\label{lammu}
\lambda=0.08(2)\,,\qquad \mu=0.21(3)\,
\end{equation}
as shown in the right panel of Fig.~\ref{FIGFS}. However, the collapse is less precise as the one for $n_0$ because of severe corrections to scaling, as mentioned before in Fig.~\ref{FIGWH}.

%
\subsection{Off-critical simulations}

\begin{figure} 
\includegraphics[width=90mm]{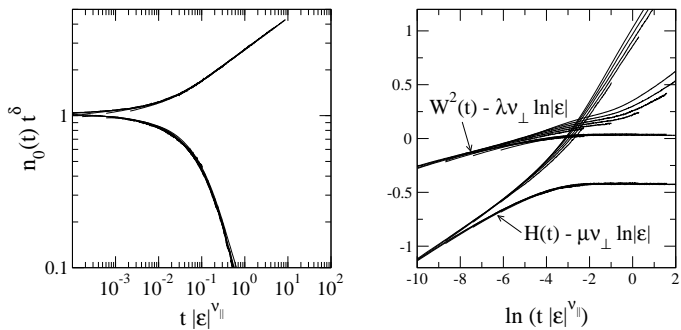}
\caption{\label{FIGOFF}
Off-critical simulations. Data collapses according to the scaling forms~(\ref{Form4})-(\ref{Form6}) for $\epsilon=\pm 0.0004,\ldots ,\pm 0.0128$. Data points for $t<25$ are eliminated in order to avoid deviations due to initial transients.
}
\end{figure}

So far we determined two ratios of the three standard exponents $\beta,\nu_\perp,\nu_\parallel$. In order to determine the third exponent one has to perform off-critical simulations. Here we expect the scaling forms~\cite{Alon2}
\begin{eqnarray}
\label{Form4}
n_0(\epsilon,t) &=& t^{-\delta} g(t |\epsilon|^{\nu_\parallel}) \\
\exp\bigl(W^2(\epsilon,t)\bigr) &=&
\label{Form5}
|\epsilon|^{-\lambda\nu_\perp} G_1(t |\epsilon|^{\nu_\parallel})\\
\label{Form6}
\exp\bigl(H(\epsilon,t)\bigr) &=&
|\epsilon|^{-\mu\nu_\perp} G_2(t |\epsilon|^{\nu_\parallel})\,,
\end{eqnarray}
where $\epsilon=q-q_c$ denotes the distance from criticality. These scaling forms should hold below and above criticality, at each case with different scaling functions. The corresponding data collapses are shown in Fig.~\ref{FIGOFF}. It turns out that the best collapses for the bottom layer density $n_0$ are obtained for
\begin{equation}
\nu_\parallel=1.85(5).
\end{equation}
In the smooth phase (lower branches) the interface width and the average height show reasonable data collapses for $\lambda \nu_\perp = 0.09(2)$ and $\mu \nu_\perp = 0.235(15)$. With $\nu_\perp\approx 1.1$ (taken from Eq.~(\ref{ChateExpon})) these estimates are compatible with Eq.~(\ref{lammu}). In the rough phase (upper branches) there is no convincing data collapse, indicating that the roughening interface crosses over to a different type of dynamic critical behavior, probably related to the KPZ equation.
%
%

\section{Conclusions}
%
To summarize we have introduced a solid-on-solid growth process which exhibits a novel type of roughening transition in 1+1 dimensions. The dynamic rules are defined in such a way that the sites at the bottom layer (the spontaneously selected lowest layer of the interface) evolve effectively as a pair contact process with diffusion $2A\to 3A$, $2A \to \emptyset$. Measruring the bottom layer density $n_0$ in finite-size and off-critical Monte Carlo simulations we have estimated the critical exponents
\begin{equation}
\delta=0.21(2)\,,\quad
z=1.75(2)\,,\quad
\nu_\parallel=1.85(5)\,.
\end{equation}
These exponents  are in good agreement with recent estimates for the PCPD reported in~\cite{Chate02}, confirming that the roughening transition is in fact driven by a pair conact process at the bottom layer. By comparing these exponents we implicitly assume that the PCPD is characterized by simple power-law scaling with well-defined critical exponents, a postulate which is still debated~\cite{Hinrichsen02,Chate02,Odor02}. But even in terms of effective exponents the numerical results clearly confirm that the roughening transition in the proposed model is in fact driven by an underlying pair contact process with diffusion.

In the active phase of the PCPD the interface is smooth an pinned to the bottom layer, while it roughens in the inactive phase. Since in the inactive phase of the PCPD the particle density decays algebraically as $n_0\sim 1/\sqrt{t}$, the interface does not propagate uniformly, instead it remains pinned and becomes increasingly facetted. At criticality the interface width and the average height of the interface grow logarithmically with time or even slower. Adopting the scaling picture outlined in Ref.~\cite{Alon2}, the exponentiated quantities $\exp[W^2(t)]$ and $\exp[H(t)]$ are expected to obey ordinary power-law scaling with corresponding critical exponents $\lambda\simeq 0.08$ and  $\mu\simeq 0.21$, respectively. However, these estimates have to be taken with care as both quantities, especially the interface width, show strong deviations from power-law scaling. 

Unlike the dimer model introduced in~\cite{Dimer}, the present model does not exhibit a facetting transition. In the PC-driven case, where dimers are deposited {\em and} evaporated, the facetting transition is related to a biased diffusion of steps along inclined parts of the interface with slope $1$. Such a step moves upward (downward) if a dimer is deposited (evaporated) so that for $q>1/2$ all terraces are swept to the top of the pyramids, stabilizing the facetted state. In the present model, where monomers evaporate instead of dimers, there is no such mechanism.

As in the case of DP- and PC-driven growth processes, the critical behavior at the first few layers above the bottom layer may be described by a unidirectionally coupled sequence of diffusive pair contact processes. In fact, the densities $n_1(t)$ and $n_2(t)$ (not shown here) display the same qualitative behavior as the particle densities in such a coupled hierarchy. However, a systematic study of a coupled sequence of PCPD's should be postponed until the PCPD itself is fully understood. 

The results of the present work suggest that any absorbing phase transition can be used generate a roughening transition in an appropriately defined growth process. For example, a generalization of the present model to a recently introduced  triplet process~\cite{Triplet} is straight forward. Moreover, it would be interesting to study the influence of evaporation from the middle of plateaus combined with a hard-core wall at zero height, as it was done in the case of DP-related growth processes in Refs.~\cite{Wetting}.

\vspace{2mm}
\noindent {\bf Acknowledgements:} The simulations were partly performed on the ALiCE parallel computer at the IAI in Wuppertal. I would like to thank B. Orth and G. Arnold for technical support.

\newpage

\end{document}